\IfFileExists{IEEEtran.cls}%
  {\documentclass[conference]{IEEEtran}}%
  {\documentclass[10pt,twocolumn,letterpaper]{article}}

\makeatletter
\newif\ifusingieee
\@ifclassloaded{IEEEtran}{\usingieeetrue}{\usingieeefalse}
\makeatother

\usepackage[T1]{fontenc}
\usepackage[utf8]{inputenc}
\usepackage{graphicx}
\usepackage{amsmath,amssymb}
\usepackage{booktabs}
\usepackage{xcolor}
\usepackage{url}
\usepackage{balance}

\ifusingieee
  \IEEEoverridecommandlockouts
  \usepackage{cite}
\else
  \usepackage{mathptmx}
  \usepackage[letterpaper,top=0.75in,bottom=1.0in,left=0.625in,
              right=0.625in,columnsep=0.2in]{geometry}
  \usepackage{caption}
  \newcommand{\IEEEauthorblockN}[1]{#1\\[2pt]}
  \newcommand{\IEEEauthorblockA}[1]{\small #1}
  \newenvironment{IEEEkeywords}%
    {\vspace{0.4em}\noindent\small\textbf{\textit{Index Terms}}---}%
    {\vspace{0.6em}}
  \usepackage{etoolbox}
  \AtBeginEnvironment{thebibliography}{%
    \footnotesize\setlength{\itemsep}{0.25ex}\setlength{\parsep}{0pt}}
\fi

\usepackage[hidelinks]{hyperref}

\title{Perception-Limited UAV Positioning: On the Reliability of
Line-of-Sight Constraints in Sensing-Driven Aerial Networks}

\author{%
\IEEEauthorblockN{Kamal Shayegan}
\IEEEauthorblockA{Telecommunication Company of Iran\\
Email: K1470sh@gmail.com}%
}
\date{}

\begin{document}
\maketitle

\begin{abstract}
Unmanned Aerial Vehicles carrying access points or base stations are
widely positioned so as to hold line-of-sight links with every ground
user at once. The algorithms that do so are evaluated against obstacle
information treated as exact, whether that information comes from a
surveyed map or, more recently, from onboard vision. This paper analyses
what becomes of that requirement when the obstacle information is
estimated rather than given, and derives the detection accuracy it
implicitly demands. Three results follow. The probability of satisfying
the requirement decays geometrically with the number of users, so that a
detector accurate enough for one user is not accurate enough for a
network. Spatially correlated detection error raises that probability but
concentrates failures into systematic blind spots. Relaxing the
requirement from all users to ninety per cent of them widens the tolerable
detection error by more than an order of magnitude, far beyond what any
plausible improvement of the detector could deliver. The requirement,
rather than perception accuracy, is the binding limitation.
\end{abstract}


\section{Introduction}
Unmanned Aerial Vehicles (UAVs) carrying Wi-Fi Access Points (APs) or
cellular Base Stations (BSs) are an established means of extending
coverage and capacity where terrestrial infrastructure is absent, damaged
or temporarily overloaded~\cite{zeng2016uav,mozaffari2019tutorial}. Their
value rests on placement: because air-to-ground links degrade sharply
without Line-of-Sight (LoS), where the UAV hovers largely determines the
Quality of Service (QoS) the aerial network can deliver.

The placement problem has been studied from several directions. An early
and influential strand treats it analytically, deriving the altitude that
maximises ground coverage from a statistical model of the urban
environment~\cite{alhourani2014optimal}, and extending the result to the
deployment of several UAVs at once~\cite{mozaffari2016efficient}. A second
strand formulates placement as a constrained optimisation over position
and transmit power, solved by convex relaxation or by search over a
discretised volume~\cite{boryaliniz2016efficient,lyu2017placement}. Both
rely on a probabilistic blockage model in which LoS availability is a
function of elevation angle and an environment class, rather than of the
particular buildings present. That abstraction is well suited to coverage
planning and poorly suited to a specific venue, where a single building
in the wrong place determines whether a given user is served. The
measurement literature supports this reading: surveys of air-to-ground
channel modelling report wide variation across environments and note that
statistical models capture the average rather than the
instance~\cite{khuwaja2018survey}.

A third strand applies learning. Deep reinforcement learning has been used
for interference management in cellular-connected
UAVs~\cite{challita2019interference} and for trajectory optimisation of
flying base stations~\cite{bayerlein2018trajectory}, building on the value
approximation methods that made deep Q-learning
practical~\cite{mnih2015human,sutton2018reinforcement}. The attraction is
that a learned policy need not assume a tractable closed form for the
environment, which is precisely what obstacle-rich urban deployment
denies.

Within that third strand, one family of algorithms is distinguished by
treating obstacles explicitly rather than probabilistically. Shafafi et
al. introduced the Traffic- and Obstacle-aware Positioning Algorithm
(TOPA)~\cite{shafafi2023topa}, later recast as a Deep Q-Network
formulation, RLTOPA~\cite{shafafi2024rltopa}, generalised into a
simulation framework, RLpos-3~\cite{shafafi2025rlpos3}, extended to
several UAVs as EMTAD~\cite{shafafi2025emtad}, and most recently given an
onboard vision front end as VTOPA~\cite{shafafi2026vtopa}. An early
statement of the problem appears in~\cite{shafafi2023dce}. Across that
progression the objective is remarkably stable: maximise aggregate
throughput subject to maintaining LoS with \emph{every} ground user
simultaneously.

This paper is concerned with the status of the information used to decide
whether that constraint holds. In~\cite{shafafi2023topa,shafafi2024rltopa,
shafafi2025rlpos3} the obstacle geometry is read from a building table
held in advance and the LoS indicator is computed from it by ray
intersection, so the constraint is evaluated against ground truth.
VTOPA~\cite{shafafi2026vtopa} replaces that table with a vision pipeline.
This is the natural direction of travel, consistent with the sensing
architectures surveyed in~\cite{shafafi2026airspace} and with the wider
move toward integrated sensing and communications~\cite{liu2022isac}, but
it changes the constraint from a statement about known geometry into a
statement about an estimate. No work in this line, to our knowledge,
quantifies what that substitution costs.

The contribution of this paper is analytical rather than algorithmic, and
is threefold. First, we isolate the perception assumption embedded in
LoS-constrained positioning, showing that obstacle geometry never enters
the observation space of the learning formulations directly, but only
through a scalar count of users in LoS that the environment supplies
exactly; the learned policy therefore optimises against an oracle it has
no means of questioning. Second, we derive in closed form the probability
that the all-user LoS constraint is correctly satisfied when the LoS
decision is a classification with non-zero error rate, first under
independent errors and then under a Beta-Binomial model that admits
spatial correlation between users sharing a blocker, and we invert both
expressions to obtain the detection accuracy a sensing subsystem must
reach for a stated reliability target. Third, we compare that requirement
against the alternative of relaxing the constraint itself, from service to
all users to service to a specified fraction of them, which turns the
analysis into a design trade between sensing specification and service
model. We also argue that Reconfigurable Intelligent Surfaces (RIS) remove
the justification for a geometric LoS constraint altogether, since a
reflected path serves a user who has no direct one, and we indicate where
a reformulated positioning function belongs within a layered sensing
architecture. No simulation is required for any of this; every result is
closed form and reproducible from the expressions given.

The remainder of the paper is organised as follows.
Section~\ref{sec:model} presents the system model, the positioning problem
and the perception model, together with the notation used throughout.
Section~\ref{sec:oracle} examines how existing LoS-constrained algorithms
treat perception, and sets out the assumptions this exposes.
Section~\ref{sec:reliability} derives the reliability of the all-user
constraint under imperfect perception, extends it to correlated error, and
compares it against a relaxed constraint.
Section~\ref{sec:implications} draws out the consequences for
sensing-driven positioning. Section~\ref{sec:concl} concludes.

\section{System Model and Problem Formulation}
\label{sec:model}

\subsection{Network model}
We consider a UAV-assisted access network deployed over a square urban
venue of side $S$ metres containing $M$ buildings, which act as the
obstacles of interest. A single UAV carries an AP and is positioned at
$P=(x,y,z)$ within an admissible deployment volume $Z_p$, bounded by
$[x^{\min},x^{\max}]$, $[y^{\min},y^{\max}]$ and $[z^{\min},z^{\max}]$,
where the altitude bounds follow from regulation and from building
heights. The UAV serves $N$ ground User Equipments (UEs), with UE$_i$
located at $P_i=(x_i,y_i,z_i)$. Table~\ref{tab:notation} collects the
notation.

Time is divided into snapshots $t_k = k\,\Delta t$, $k \in \mathbb{N}_0$,
where the update period $\Delta t$ is long relative to the packet
timescale, so that UE positions and traffic demands may be treated as
constant within a snapshot. UE$_i$ demands a bit rate $B_i(t_k)$, which
requires a link capacity $C_i(t_k) \geq B_i(t_k)$, with the aggregate
bounded by the maximum channel capacity $C^{\mathrm{MAX}}$ since the
medium is shared. The demand determines the Modulation and Coding Scheme
$\mathrm{MCS}_i$ selected for the link, which in turn fixes a minimum
required Signal-to-Noise Ratio $\mathrm{SNR}_i^{\min}$. Path loss follows
the ITU-R P.1411 models for short-range outdoor
links~\cite{itu1411}, parameterised by the environment class of the venue.

Blockage is represented by a binary indicator $\mathrm{LoS}_i \in
\{0,1\}$, equal to one when the segment from $P_i$ to $P$ intersects no
building, and we write $n_{\mathrm{LoS}}(t_k) = \sum_{i}\mathrm{LoS}_i$
for the number of users in LoS at $t_k$.

\begin{table}[t]
\centering
\caption{Notation.}
\label{tab:notation}
\small
\begin{tabular}{@{}ll@{}}
\toprule
\textbf{Symbol} & \textbf{Meaning} \\
\midrule
$S$, $M$ & Venue side length (m); number of buildings \\
$N$ & Number of served UEs \\
$P$, $P_i$ & UAV position; position of UE$_i$ \\
$Z_p$ & Admissible UAV deployment volume \\
$t_k$, $\Delta t$ & Snapshot index; position update period \\
$B_i(t_k)$ & Traffic demand of UE$_i$ (bit/s) \\
$C_i(t_k)$, $C^{\mathrm{MAX}}$ & Link capacity; maximum channel capacity \\
$R_i(t_k)$ & Bit rate received from UE$_i$ \\
$\mathrm{MCS}_i$, $\mathrm{SNR}_i^{\min}$ & Coding scheme; required SNR \\
$\mathrm{LoS}_i$, $n_{\mathrm{LoS}}$ & LoS indicator; number of UEs in LoS \\
\midrule
$p$ & Per-UE LoS misclassification probability \\
$\rho$ & Intra-class correlation of those errors \\
$\eta$, $K$ & Required served fraction; $K=\lceil \eta N \rceil$ \\
$\gamma$ & Target constraint-satisfaction probability \\
\bottomrule
\end{tabular}
\end{table}

\subsection{Positioning problem}
The positioning problem places the UAV so as to maximise the aggregate
bit rate received from the UEs, subject to capacity, deployment-volume and
LoS constraints:
\begin{align}
\max_{(x,y,z)} \quad & \sum_{i=1}^{N} R_i(t_k) \label{eq:obj}\\
\text{s.t.}\quad
& \textstyle\sum_{i=1}^{N} C_i(t_k) \leq C^{\mathrm{MAX}}, \label{eq:cap}\\
& 0 < B_i(t_k) \leq C_i(t_k), \quad i=1,\dots,N, \label{eq:demand}\\
& (x,y,z) \in Z_p, \label{eq:vol}\\
& \textstyle\sum_{i=1}^{N} \mathrm{LoS}_i = N. \label{eq:constraint}
\end{align}
Constraint~\eqref{eq:constraint} is the object of this paper. It is hard
and collective: a candidate position is admissible only if every user is
in LoS simultaneously.

\subsection{Perception model}
In the formulations of Section~\ref{sec:oracle}, $\mathrm{LoS}_i$ is
computed geometrically from a building set known in advance, and is
therefore exact by construction. We replace that assumption with a
perception subsystem that returns an estimate
$\widehat{\mathrm{LoS}}_i$ obtained from sensing, whether by vision,
radar or radio measurement. Let
\begin{equation}
p \;=\; \Pr\{\widehat{\mathrm{LoS}}_i \neq \mathrm{LoS}_i\}
\label{eq:perr}
\end{equation}
be the per-user misclassification probability, assumed identical across
users. Errors on different users need not be independent, since users
close together are occluded by the same structures; we characterise their
dependence by the intra-class correlation $\rho \in [0,1)$, with $\rho=0$
recovering independence. Finally, let $\gamma$ denote the probability with
which the system is required to satisfy its LoS constraint, and $\eta$ the
fraction of users that must be served when the all-user requirement is
relaxed, giving $K=\lceil \eta N \rceil$. Throughout we assume the
remaining geometry is exact, which isolates the effect of interest and
yields an optimistic bound, since UE position error would compound it.

\section{Perception in Existing Positioning Algorithms}
\label{sec:oracle}
Table~\ref{tab:algorithms} summarises the algorithms that solve
\eqref{eq:obj}--\eqref{eq:constraint} or close variants of it.
TOPA~\cite{shafafi2023topa} positions a single UAV geometrically.
RLTOPA~\cite{shafafi2024rltopa} recasts the problem as a Markov Decision
Process solved with a Deep Q-Network, reporting gains of up to $95\%$ in
aggregate throughput and $71\%$ in mean delay over a geometric baseline.
RLpos-3~\cite{shafafi2025rlpos3} is a framework rather than an algorithm,
coupling TensorFlow Agents and OpenAI Gym to Network Simulator 3 through
the ns3-gym bridge~\cite{gawlowicz2019ns3gym} so that the observation
space, action space and reward can be redefined per scenario.
EMTAD~\cite{shafafi2025emtad} lifts the single-UAV restriction. VTOPA
\cite{shafafi2026vtopa} removes the dependence on a surveyed map by
extracting obstacle information from onboard vision.

\begin{table*}[t]
\centering
\caption{LoS-constrained UAV positioning algorithms. Every entry treats
LoS as a binary quantity required to hold for all served users
simultaneously.}
\label{tab:algorithms}
\small
\begin{tabular}{@{}lllllp{0.30\textwidth}@{}}
\toprule
\textbf{Algorithm} & \textbf{Ref.} & \textbf{Obstacle knowledge} &
\textbf{UAVs} & \textbf{Method} & \textbf{Reported outcome} \\
\midrule
TOPA    & \cite{shafafi2023topa}   & Surveyed map, static & Single &
Geometric / heuristic & 100\% improvement in the aggregate throughput \\
RLTOPA  & \cite{shafafi2024rltopa} & Surveyed map, static & Single &
Deep Q-Network & Up to $95\%$ aggregate throughput gain, $71\%$ mean
delay reduction \\
RLpos-3 & \cite{shafafi2025rlpos3} & Surveyed map, static & Single &
Framework over ns-3 & Reproducible benchmarking; no algorithmic claim \\
EMTAD   & \cite{shafafi2025emtad}  & Surveyed map, static & Multiple &
Scalable heuristic & Reducing
UAV requirements by up to 67\% and enhancing aggregate
throughput by 400\% \\
VTOPA   & \cite{shafafi2026vtopa}  & Onboard vision & Single &
Vision-aided positioning & up to 50\% increase in aggregate throughput and 50\% reduction in delay\\
\bottomrule
\end{tabular}
\end{table*}

The learning formulations share an observation space of the form
\begin{equation}
\mathrm{obs}[t_k] = \big\{ \{P_i\}_{i=1}^{N},\; P(t_k),\;
n_{\mathrm{LoS}}(t_k) \big\},
\label{eq:obs}
\end{equation}

the user coordinates, the UAV coordinates, and the number of users
currently in LoS. Obstacles are absent from~\eqref{eq:obs}. The building
geometry reaches the decision process only through the scalar
$n_{\mathrm{LoS}}(t_k)$, computed by ray intersection against a table the
processing system holds in advance. The agent therefore does not learn to
perceive or avoid obstacles; it learns to ascend a gradient in a quantity
supplied exactly and instantaneously by what is, in effect, an oracle. The
physical environment is compressed into one integer per snapshot, and that
integer is assumed correct.

VTOPA~\cite{shafafi2026vtopa} is significant precisely because it attacks
this assumption, deriving obstacle information from vision rather than
from a survey. Doing so changes the status of $n_{\mathrm{LoS}}$ from a
measured fact to an estimate, while~\eqref{eq:constraint} remains a hard
constraint on exactly that estimate. Several further assumptions become
visible at the same time. Obstacles are modelled as static buildings,
though vehicles and crowds move within $\Delta t$. UE positions are taken
as exact, so their error would propagate into the LoS geometry. And the
link is treated as insensitive to UAV heading, although experimental
characterisation of air-to-ground and ground-to-air channels by the same
authors shows that heading and the resulting antenna radiation pattern
measurably affect received signal strength and
throughput~\cite{shafafi2024a2g}. We analyse the first of these, which is
the one that bears directly on~\eqref{eq:constraint}.

\section{Reliability of the All-User LoS Constraint}
\label{sec:reliability}

\subsection{Independent classification error}
\begin{figure}[t]
\centering
\includegraphics[width=\columnwidth]{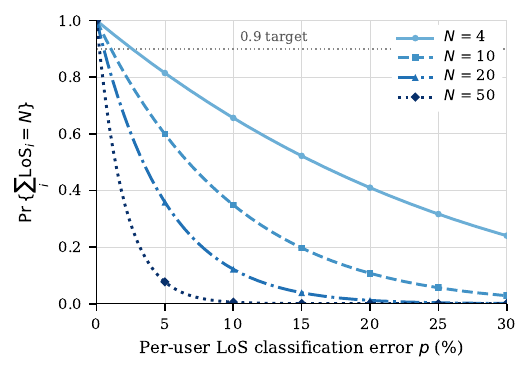}
\caption{Probability that the all-user LoS constraint
\eqref{eq:constraint} holds under independent per-user classification
error.}
\label{fig:indep}
\end{figure}

Under the perception model of Section~\ref{sec:model} with $\rho=0$, each
user is classified independently with error probability $p$, and
\begin{equation}
\Pr\Big\{ \textstyle\sum_{i=1}^{N}\mathrm{LoS}_i = N \Big\} = (1-p)^{N}.
\label{eq:indep}
\end{equation}
Fig.~\ref{fig:indep} plots~\eqref{eq:indep}. The decay is geometric in
network size rather than graceful. A detector with $5\%$ per-user error,
unremarkable in isolation, satisfies the constraint in $60\%$ of snapshots
at $N=10$ and $36\%$ at $N=20$. At $10\%$ error these fall to $35\%$ and
$12\%$ respectively, and at $N=50$ the constraint is essentially never
satisfied.

Inverting~\eqref{eq:indep}, the accuracy required for the constraint to
hold with probability at least $\gamma$ is
\begin{equation}
p \;\le\; 1 - \gamma^{1/N}.
\label{eq:requirement}
\end{equation}
For $\gamma=0.9$ this is $2.60\%$ at $N=4$, $1.05\%$ at $N=10$, $0.53\%$
at $N=20$ and $0.21\%$ at $N=50$. Formulated as it stands, the problem
implicitly demands sub-$1\%$ LoS classification error for networks of
even modest size.

\subsection{Spatially correlated error}
Independence is a convenience. Users close together share blockers, so
misclassifications are positively correlated in practice. We model this
with a Beta-Binomial: let the per-snapshot error rate be a random variable
$Q \sim \mathrm{Beta}(a,b)$ with $\mathbb{E}[Q]=p$, and let users be
conditionally independent given $Q$. Writing $s=a+b$, the intra-class
correlation is $\rho = 1/(s+1)$, so that $s=(1-\rho)/\rho$, $a=ps$ and
$b=(1-p)s$. The probability of no misclassification becomes
\begin{equation}
\Pr\Big\{ \textstyle\sum_{i}\mathrm{LoS}_i = N \Big\} =
\frac{\Gamma(b+N)\,\Gamma(s)}{\Gamma(b)\,\Gamma(s+N)} ,
\label{eq:corr}
\end{equation}
which recovers $(1-p)^N$ as $\rho \to 0$ and tends to $1-p$ as
$\rho \to 1$.

\begin{figure}[t]
\centering
\includegraphics[width=\columnwidth]{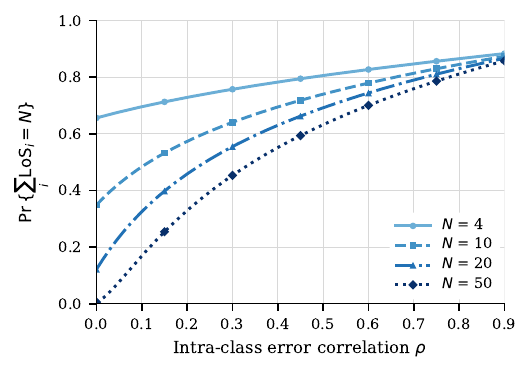}
\caption{Effect of spatially correlated classification error on the
all-user constraint, at a fixed marginal error $p=10\%$. Independence
($\rho=0$) is the worst case.}
\label{fig:corr}
\end{figure}

Fig.~\ref{fig:corr} plots~\eqref{eq:corr} at a fixed marginal error of
$10\%$. Correlation is strongly favourable to the all-user constraint: at
$N=50$ the satisfaction probability rises from $0.5\%$ under independence
to $45\%$ at $\rho=0.3$. Fig.~\ref{fig:indep} should therefore be read as
a worst case at fixed marginal error, not as a prediction.

This favourable reading is incomplete. The mean number of
misclassifications is $Np$ regardless of $\rho$, but its variance is
\begin{equation}
\mathrm{Var} = N p (1-p)\big[1 + (N-1)\rho\big],
\label{eq:var}
\end{equation}
an inflation of $6.7\times$ at $N=20$, $\rho=0.3$ and $15.7\times$ at
$N=50$. Correlation does not reduce error; it concentrates it. Snapshots
are either clean or badly wrong, and the bad ones correspond to systematic
blind spots, typically a blocker occluding a group of users at once. For
an all-or-nothing constraint this is benign. For any service guarantee
expressed over users it is worse than independent error, because the
failures arrive together.

\subsection{Relaxing the constraint}
The preceding results suggest that the binding problem is the constraint
rather than the detector. Replace~\eqref{eq:constraint} by a requirement
to serve a fraction $\eta$ of the users, so that at least $K=\lceil \eta N
\rceil$ must be in LoS. Under independent error the number of correctly
served users is Binomial$(N,1-p)$ and
\begin{equation}
\Pr\Big\{ \textstyle\sum_{i}\mathrm{LoS}_i \geq K \Big\} =
\sum_{j=K}^{N} \binom{N}{j} (1-p)^{j} p^{\,N-j}.
\label{eq:relaxed}
\end{equation}

\begin{figure}[t]
\centering
\includegraphics[width=\columnwidth]{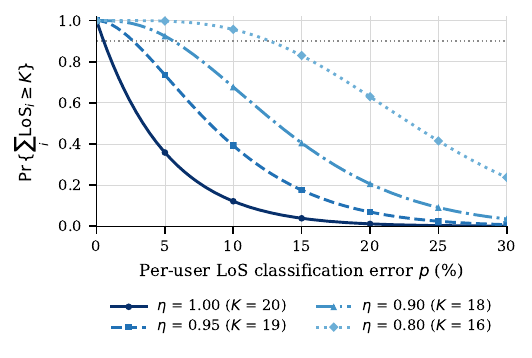}
\caption{Relaxing the constraint from all users to a served fraction
$\eta$, for $N=20$ under independent error.}
\label{fig:relaxed}
\end{figure}

Fig.~\ref{fig:relaxed} plots~\eqref{eq:relaxed} for $N=20$. The effect is
large. At $p=10\%$ the satisfaction probability rises from $12\%$ at
$\eta=1$ to $68\%$ at $\eta=0.9$ and $96\%$ at $\eta=0.8$.
Table~\ref{tab:requirements} states the same result as a design
requirement: at $N=20$ and $\gamma=0.9$, admitting one unserved user in
ten widens the tolerable classification error from $0.53\%$ to $5.64\%$,
better than a tenfold relaxation of the sensing specification. At $N=4$
the two columns coincide, because a $90\%$ fraction of four users rounds
up to four.

\begin{table}[t]
\centering
\caption{Maximum tolerable per-user LoS classification error for a $90\%$
constraint-satisfaction target.}
\label{tab:requirements}
\small
\begin{tabular}{@{}cccc@{}}
\toprule
$N$ & All users, $\rho=0$ & All users, $\rho=0.3$ &
$\eta=0.9$, $\rho=0$ \\
\midrule
4  & $2.60\%$ & $3.88\%$ & $2.60\%$ \\
10 & $1.05\%$ & $2.42\%$ & $5.45\%$ \\
20 & $0.53\%$ & $1.82\%$ & $5.64\%$ \\
50 & $0.21\%$ & $1.35\%$ & $6.43\%$ \\
\bottomrule
\end{tabular}
\end{table}

The practical reading is that effort spent relaxing the service model buys
more than effort spent improving perception, and that the two are not
interchangeable: no achievable detector accuracy rescues an all-user
constraint at $N=50$, whereas a modest relaxation does.

\section{Implications for Sensing-Driven Positioning}
\label{sec:implications}

\subsection{From an MDP to a POMDP}
Once $n_{\mathrm{LoS}}$ is estimated, the agent in~\eqref{eq:obs} observes
a noisy function of the state rather than the state, and the problem is
properly a Partially Observable Markov Decision
Process~\cite{kaelbling1998planning}. A policy trained against an exact
indicator has no representation of its own uncertainty and no way to act
on it. The difficulty is compounded by the reward
in~\cite{shafafi2024rltopa}, which is built from the same quantity:
estimation error enters the training signal directly, so the agent
optimises a noisy and potentially biased objective rather than merely
acting on a noisy input. RLpos-3~\cite{shafafi2025rlpos3} is well placed
to study this, since its observation space and reward are configurable; a
perturbation layer injecting controlled error into the LoS indicator would
turn the analysis above into a measured degradation curve.

\subsection{Sensing as an action}
Under integrated sensing and communications~\cite{liu2022isac}, sensing
consumes time and spectrum that could otherwise carry traffic. Choosing
when and where to sense therefore becomes a decision rather than a free
background service, and the sensing cadence sets the snapshot period
$\Delta t$ against which obstacle motion must be judged. The action space
in~\cite{shafafi2024rltopa}, six grid moves and a stay, cannot express
this. Where the perception result must be computed off-board, the
reliability of the link carrying it becomes part of the problem, a concern
with a long history in error-resilient source
coding~\cite{shafafi2009reversible}.

\subsection{RIS and the end of binary LoS}
A more fundamental difficulty arises with RIS. A reconfigurable surface
mounted on a building creates a controllable reflected
path~\cite{wu2019intelligent,direnzo2020smart}, so a user with no
geometric LoS to the UAV may nonetheless be served at the required rate.
The indicator $\mathrm{LoS}_i$ then no longer describes what the network
can deliver, and~\eqref{eq:constraint}, a statement about geometry, stops
being a proxy for QoS. Improving the detector does not repair this: the
objective is measuring the wrong thing.

The repair is to express the constraint over served rate rather than
geometry. The machinery is already present, since $\mathrm{MCS}_i$ is
selected from $B_i(t_k)$ and the formulation therefore reasons in terms of
$\mathrm{SNR}_i^{\min}$. Replacing~\eqref{eq:constraint} with a per-user
rate condition evaluated over the effective channel, direct or
RIS-assisted, would make the formulation agnostic to how the link is
realised, and would subsume the relaxation of
Section~\ref{sec:reliability} as a special case.

\subsection{Architectural placement}
The AIRSPACE reference architecture~\cite{shafafi2026airspace} indicates
where these functions belong. Its perception layer supplies what the
positioning formulations assume: sensing data, channel state information
and target detection. Its edge intelligence layer is the natural home for
the positioning policy, since the fast inner loop between perception and
edge operates on the timescale at which placement and RIS configuration
must adapt, while mission policy remains in the slower outer loop.
Positioning is thus better understood as one control loop inside a sensing
architecture than as a standalone optimisation fed by a static map.
Training such a policy need not rely on gradient methods alone;
population-based optimisation of neural networks has proven effective in
other signal-classification settings~\cite{pourvahab2025pel}, and is a
candidate where the reward is noisy by construction.

\section{Conclusions}
\label{sec:concl}
LoS-constrained UAV positioning algorithms are evaluated against a
perception layer treated as exact. This paper quantified what happens when
it is not. Under independent per-user classification error the all-user
LoS constraint holds with probability $(1-p)^N$, which demands sub-$1\%$
accuracy at $N=10$ for a $90\%$ reliability target. Spatial correlation
raises that probability but inflates the variance of the number of served
users by $1+(N-1)\rho$, concentrating failures into systematic blind
spots. Relaxing the constraint from all users to $90\%$ of them widens the
tolerable error by more than an order of magnitude at $N=20$, beyond what
any plausible improvement in detection could achieve.

The conclusion we draw is that the all-user LoS constraint, rather than
perception accuracy, is the binding limitation on sensing-driven UAV
positioning, and that RIS removes its justification entirely by making
geometric LoS a poor proxy for deliverable rate. Reformulating the
constraint over served rate, and treating positioning as a partially
observable problem inside a sensing architecture, appears more promising
than further refinement of the detector.



\bibliographystyle{IEEEtran}
\bibliography{IEEEabrv,Ref}

\end{document}